\documentclass[a4paper,twoside,prd,nofootinbib,preprintnumbers,twocolumn]{revtex4}
\usepackage{amssymb,amsmath,bm,natbib}
\usepackage{color}
\usepackage{slashed}
\usepackage{graphics}
\usepackage{graphicx}
\usepackage[utf8]{inputenc}
\usepackage[caption=false]{subfig}
\usepackage{hyperref}
\usepackage{url}
\usepackage{dsfont}
\usepackage{float} 
\usepackage{cancel}

\newcommand{\eq}[1]{Eq.~\eqref{#1}}

\begin{document}
\preprint{CERN-TH-2020-177, PSI-PR-20-18,   UZ-TH  40/20}
\title{Combined Explanation of the $Z\to b\bar b$ Forward-Backward Asymmetry, \\the Cabibbo Angle Anomaly, $\tau\to\mu\nu\nu$ and $b\to s\ell^+\ell^-$ Data}

\author{Andreas Crivellin}
\email{andreas.crivellin@cern.ch}
\affiliation{CERN Theory Division, CH--1211 Geneva 23, Switzerland}
\affiliation{Paul Scherrer Institut, CH--5232 Villigen PSI, Switzerland}
\affiliation{Physik-Institut, Universit\"at Z\"urich, Winterthurerstrasse 190, CH--8057 Z\"urich, Switzerland}

\author{Claudio Andrea Manzari}
\email{claudioandrea.manzari@physik.uzh.ch}
\affiliation{Paul Scherrer Institut, CH--5232 Villigen PSI, Switzerland}
\affiliation{Physik-Institut, Universit\"at Z\"urich, Winterthurerstrasse 190, CH--8057 Z\"urich, Switzerland}

\author{Marcel Alguer\'{o}}
\email{malguero@ifae.es}
\affiliation{Grup de Fisica Te\`orica (Departament de Fisica), Universitat Aut\`onoma de Barcelona, E-08193 Bellaterra (Barcelona), Spain.}
\affiliation{Institut de Fisica d'Altes Energies (IFAE),
	The Barcelona Institute of Science and Technology, Campus UAB, E-08193 Bellaterra (Barcelona), Spain.}

\author{Joaquim Matias}
\email{matias@ifae.es}
\affiliation{Grup de Fisica   Te\`orica (Departament de Fisica), Universitat Aut\`onoma de Barcelona, E-08193 Bellaterra (Barcelona), Spain.}
\affiliation{Institut de Fisica d'Altes Energies (IFAE),
	The Barcelona Institute of Science and Technology, Campus UAB, E-08193 Bellaterra (Barcelona), Spain.}

\begin{abstract}
In this article we propose a simple model which can provide a combined explanation of the $Z\to b\bar b$ forward-backward asymmetry, the Cabibbo Angle Anomaly (CAA), $\tau\to\mu\nu\nu$ and $b\to s\ell^+\ell^-$ data. This model is obtained by extending the Standard Model (SM) by two heavy vector-like quarks (an $SU(2)_L$ doublet (singlet) with hypercharge $-5/6$ (-1/3)), two new scalars (a neutral and a singly charged one) and a gauged $L_\mu-L_\tau$ symmetry. The mixing of the new quarks with the SM ones, after electroweak symmetry breaking, does not only explain $Z\to b\bar b$ data but also generates a lepton flavour universal  contribution to $b\to s\ell^+\ell^-$ transitions. Together with the lepton flavour universality violating effect, generated by loop-induced $Z^\prime$ penguins involving the charged scalar and the heavy quarks, it gives an excellent fit to data ($6.1\,\sigma$ better than the SM). Furthermore, the charged scalar (neutral vector) gives a necessarily constructive tree-level (loop) effect in $\mu\to e\nu\nu$ ($\tau\to \mu\nu\nu$), which can naturally account for the CAA (${\rm Br}[\tau\to\mu\nu\nu]/{\rm Br}[\tau\to e\nu\nu]$ and ${\rm Br}[\tau\to\mu\nu\nu]/{\rm Br}[\mu\to e\nu\nu]$).
\end{abstract}
\maketitle

\newpage
\section{Introduction}

The Standard Model (SM) of particle physics has been very successfully tested with great precision in the last decades. Nonetheless, and despite the fact that the LHC has not observed any additional particle directly, it is clear that the SM cannot be the ultimate fundamental theory of physics. In particular, it has to be extended to account for Dark Matter and Neutrino masses, but neither the scale nor the concrete nature of the additional particles is unambiguously established. Fortunately, in the flavour sector we obtained intriguing (indirect) hints for physics beyond the SM at the (multi) TeV scale. 

In particular, global fits~\cite{Capdevila:2017bsm,Altmannshofer:2017yso,DAmico:2017mtc,Ciuchini:2017mik,Hiller:2017bzc,Geng:2017svp,Hurth:2017hxg,Alok:2017sui,Alguero:2019ptt,Aebischer:2019mlg,Ciuchini:2019usw} to $b\to s\ell^+\ell^-$ data~\cite{Aaij:2014pli,Aaij:2014ora,Aaij:2015esa,Aaij:2015oid,Khachatryan:2015isa,ATLAS:2017dlm,CMS:2017ivg,Aaij:2017vbb} point convincingly towards new physics (NP) and several simple model-independent scenarios are significantly preferred over the SM hypothesis (by more then $5\,\sigma$). Furthermore, many NP models were proposed that give rise to these scenarios; including $Z^\prime$ models~\cite{Buras:2013qja,Gauld:2013qba,Gauld:2013qja,Altmannshofer:2014cfa,Crivellin:2015mga,Crivellin:2015lwa,Niehoff:2015bfa,Carmona:2015ena,Falkowski:2015zwa,Celis:2015eqs,Celis:2015ara,Crivellin:2015era,Chiang:2016qov,Crivellin:2016ejn,GarciaGarcia:2016nvr,Altmannshofer:2016oaq,Faisel:2017glo,King:2017anf,Chiang:2017hlj,DiChiara:2017cjq,Ko:2017lzd,Sannino:2017utc,Falkowski:2018dsl,Benavides:2018rgh,Maji:2018gvz,Singirala:2018mio,Guadagnoli:2018ojc,Allanach:2018lvl,Duan:2018akc,King:2018fcg,Kohda:2018xbc,Dwivedi:2019uqd,Foldenauer:2019vgn,Ko:2019tts,Allanach:2019iiy,Altmannshofer:2019xda,Calibbi:2019lvs,Aebischer:2019blw,deAnda:2020hcf,Kawamura:2019rth}, leptoquarks~\cite{Hiller:2016kry,Barbieri:2015yvd,Barbieri:2016las,Crivellin:2017dsk,Crivellin:2018yvo,Bordone:2018nbg,Crivellin:2019szf,Bordone:2019uzc,Bernigaud:2019bfy,Aebischer:2018acj,Fuentes-Martin:2019ign,Fajfer:2015ycq,deMedeirosVarzielas:2019lgb,Varzielas:2015iva,Saad:2020ihm,Saad:2020ucl,Gherardi:2020qhc,Bordone:2020lnb} or loop effects involving top-quarks~\cite{Aebischer:2015fzz,Kamenik:2017tnu} or new scalars and fermions~\cite{Gripaios:2015gra,Arnan:2016cpy,Grinstein:2018fgb,Arnan:2019uhr}. While the large majority of these models generates simple patterns which are mostly purely Lepton Flavour Universality (LFU) violating
, it has been shown in Refs.~\cite{Alguero:2018nvb,Alguero:2019ptt} that slightly more elaborated patterns with both LFU and LFU violating effects can describe data even better.

Furthermore, the anomaly in $b\to s\ell^+\ell^-$ data is accompanied by additional hints for the violation of LFU which are very interesting, even though they are statistically less significant: i) $R(D^{(*)})$~\cite{Lees:2012xj,Lees:2013uzd,Aaij:2015yra,Aaij:2017deq,Aaij:2017uff,Abdesselam:2019dgh} points towards $\tau-\mu$ LFU violation with a significance of $>\!\!3\,\sigma$~\cite{Amhis:2016xyh,Murgui:2019czp,Shi:2019gxi,Blanke:2019qrx,Kumbhakar:2019avh} ii) the anomalous magnetic moment 
of the muon 
$a_\mu$~\cite{Bennett:2006fi}
prefers NP coupling to muons by $3.7\,\sigma$~\cite{Aoyama:2020ynm} iii) ${\rm Br}[\tau\to\mu\nu\nu]/{\rm Br}[\tau\to e\nu\nu]$ and ${\rm Br}[\tau\to\mu\nu\nu]/{\rm Br}[\mu\to e\nu\nu]$ are indications for LFU violation with a significance of $\approx2\,\sigma$ iv) the deficit in $1^{\rm st}$ row CKM unitarity, known as the Cabibbo Angle Anomaly (CAA), is at the $\approx2-4\,\sigma$ level~\cite{Belfatto:2019swo,Grossman:2019bzp,Seng:2020wjq} and can possibly be interpreted as a sign of LFU violation~\cite{Coutinho:2019aiy,Crivellin:2020lzu,Crivellin:2020ebi,Coutinho:2020xhc,Kirk:2020wdk,Alok:2020jod}.

Interestingly, it has been shown that NP models can provide combined explanations of these anomalies together with $b\to s\ell^+\ell^-$ data. For example, common solutions of the $b\to s\ell^+\ell^-$ anomalies together with $a_\mu$~\cite{Bauer:2015knc,Belanger:2015nma,Crivellin:2019dun,Crivellin:2019dwb,Altmannshofer:2020axr,Huang:2020ris} and/or $b\to c\tau\nu$ data~\cite{Alonso:2015sja,Bhattacharya:2014wla,Calibbi:2015kma,Bhattacharya:2016mcc,Buttazzo:2017ixm,DiLuzio:2017vat,Calibbi:2017qbu,Blanke:2018sro,Kumar:2018kmr,Crivellin:2018yvo,Cornella:2019hct,Babu:2020hun} were studied, mostly within leptoquark models and also the CAA was correlated to $b\to s\ell^+\ell^-$ data using a heavy vector boson in the adjoint representations of $SU(2)_L$~\cite{Capdevila:2020rrl}. 

In addition to the anomalies i)-iv), related to $b\to s\ell^+\ell^-$ in the context of LFU violation, there is also the long-standing discrepancy between the SM prediction and the LEP measurement of the $Z\to b\bar b$ forward-backward asymmetry~\cite{Abdallah:2004nh}. Here, the global fit to $Zbb$ couplings reveals a tension of $\approx 2\,\sigma$ both in the left-handed and in the right-handed coupling with a strong correlation~\cite{deBlas:2016ojx,deBlas:2016nqo}. Interestingly, even though this observable could obviously be related to $b\to s\ell^+\ell^-$ transitions via NP coupling to the bottom quark, models providing such a connection have received surprisingly little attention in the literature so far.

In this article, we want to fill this gap by presenting a model that cannot only provide a common explanation of $b\to s\ell^+\ell^-$ data and the $Z\to b\bar b$ forward-backward asymmetry but also account for $\tau\to\mu\nu\nu$ and the 
Cabibbo Angle Anomaly. Notice that  a quite large effect in $Z\to b\bar b$  w.r.t the SM is necessary (in particular in the right-handed $Zbb$ coupling), such that loop effects are in general too small to account for it. Therefore, two possibilities remain to construct a viable model: the mixing of the SM $Z$ with a neutral $Z^\prime$ boson coupling to $b\bar b$, or vector-like quarks (VLQs) mixing with the SM ones after electroweak (EW) symmetry breaking. While the former case can in fact account for $Z\to b\bar b$, it is difficult to explain simultaneously $b\to s\ell^+\ell^-$ data since the effect in $Z\mu\mu$ would be too large~\cite{Calibbi:2020dyg}. Concerning 
vector-like quarks, there is only one representation each that gives the appropriate effect in the left-handed or right-handed coupling\footnote{There is a second $SU(2)_L$ doublet VLQ which contributes with opposite sign to the right-handed $Zb\bar b$ couplings. This VLQ could only account for the anomaly via an over-compensation in the fit which is in conflict with other EW data and $b\to s\gamma$~\cite{Batell:2012ca}.}. Clearly, if these 
vector-like quarks mix with the strange quark as well, a modified $Zsb$ coupling is generated. While such modified $Z$ couplings can improve the fit in $b\to s\ell^+\ell^-$, they cannot explain the hints for LFU violation in $R(K^{(*)})$ and additional ingredients are required to fully account for all data. Therefore, we will add two new scalar (one charged and one neutral) to our particle content and extend the gauge group by a $L_\mu-L_\tau$ symmetry to obtain a LFU violating effect. Interestingly, the charged scalar turns out to have just the right quantum numbers to explain at the same time the 
Cabibbo Angle Anomaly via an effect in the determination of the Fermi constant from muon decay, while the $Z^\prime$ of the gauged $L_\mu-L_\tau$ symmetry improves the agreement with data in ${\rm Br}[\tau\to\mu\nu\nu]/{\rm Br}[\tau\to e\nu\nu]$ and ${\rm Br}[\tau\to\mu\nu\nu]/{\rm Br}[\mu\to e\nu\nu]$. 

\section{The Model}
\label{model}

Our starting point is $Z\to b \bar b$ which, as outlined in the introduction, can only be explained by adding two new heavy quarks to the SM particle content, an $SU(2)_L$ doublet with hypercharge $-5/6$ ($Q$) and an $SU(2)_L$ singlet with hypercharge $-1/3$ ($D$). They couple to right-handed down-type quarks and left-handed quark doublets, respectively, via the SM Higgs doublet ($H$) and therefore can mix with $b$ quarks after EW symmetry breaking. Due to the stringent LHC bounds on new quarks their mass must be larger than $\approx\!1$TeV~\cite{Aaboud:2018saj,CMS-PAS-B2G-17-018}. Therefore, the new quarks must be vector-like under the SM gauge group such that their masses are not confined to the EW scale. However, we assume them to be chiral under a new $U(1)^\prime$ gauge group (with coupling constant $g^\prime$) such that they cannot be arbitrarily heavy but have masses of the order of the $U(1)^\prime$ breaking scale. In fact, we charge $Q_R$ and $D_L$ under $U(1)^\prime$ while $Q_L$ and $D_R$ are neutral. 
This does not only allow for the desired mixing with the SM down-type quarks but also turns out to be crucial for generating a $Z^\prime bs$ coupling later on. All SM particles are neutral under the gauged $U(1)^\prime$ except for leptons for which we assume a $L_\mu-L_\tau$ symmetry~\cite{He:1990pn,Foot:1990mn,He:1991qd}. This symmetry cannot only naturally generate the observed pattern from the PMNS matrix~\cite{Binetruy:1996cs,Bell:2000vh,Choubey:2004hn} but also avoids stringent LEP bounds on 4-lepton contact interactions~\cite{Schael:2013ita}. In addition, we introduce two $SU(3)_c\times\,SU(2)_L$ singlet scalars with $L_\mu-L_\tau$ charge of $-1$: one electrically neutral ($S$) and the other with charge $+1$ ($\phi^+$). 
In summary, our particle content is given in Table~\ref{table1}.

\begin{table}
	\begin{tabular}{c |c c c c c c | c c c c c c}
		& $q_L$ & $d_R$ & $u_R$ & $H$ & $\ell_L$ & $e_R$ & ${{Q_L}}$&${{Q_R}}$&${{D_L}}$&${{D_R}}$&${{\phi ^ + }}$&$S$\\
		\hline
		${SU{{\left( 3 \right)}_c}}$ &3&3&3&1&1&1& 3&3&3&3&1&1\\
		${SU{{\left( 2 \right)}_L}}$&2&1&1&2&2&1& 2&2&1&1&1&1\\
		${U{{\left( 1 \right)}_Y}}$&$\frac{1}{6}$&$\frac{- 1}{3}$&$\frac{2}{3}$&$\frac{1}{2}$&${ \frac{- 1}{2}}$&$-1$& ${ \frac{-5}{6}}$&${ \frac{- 5}{6}}$& $\frac{- 1}{3}$& ${ \frac{- 1}{3}}$&1&0\\
		${U{{\left( 1 \right)}^\prime }}$&0&0&0&0&\multicolumn{2}{c}{$(0,1,-1)$}& 0&1&1&0&{ $-1$}&{ $- 1$}
	\end{tabular}
	\caption{SM particle content and the scalar and fermion fields added to it  in our model together with their representations under the gauge group $SU(3)\!\cdot\! SU(2)_L\!\cdot\!U(1)_Y\!\cdot\!U(1)^\prime$. Here the bracket $(0,1,-1)$ in the columns for left and right-handed leptons means that we assume a $L_\mu-L_\tau$ flavour symmetry.}\label{table1}
\end{table}

This allows for the following Yukawa-type interactions involving quarks
\begin{align}
- {\cal L}_Y^{Q} &= \left( {Y_{f}^d\delta_{fi}{{\bar q}_{Lf}} {d_{Ri}} + \lambda _f^D{{\bar q_{Lf}}} {D_R}} \right)\!H + \kappa _f  {{\bar q}_{Lf}} {Q_R}{\phi ^ + }\nonumber\\
&+ \left( {{\kappa_{RL}}{{\bar Q}_R}{D_L} + {\kappa_{LR}}{{\bar Q}_L}{D_R} + \lambda _i^Q {{\bar Q}_L}{d_{Ri}}} \right)\!\tilde H\label{Yukawa}\\
&+ \left( {\eta _D}{{\bar D}_L}{D_R} + {\eta _Q}{{\bar Q}_R}{Q_L} \right)\!{S^\dag }+Y_{fi}^u{{\bar q}_{Lf}}{u_{Ri}}\tilde H +{\rm h.c.}  \nonumber \,,
\end{align}
where $\tilde H=i \sigma_2H^*$ is the complex conjugate of the SM Higgs doublet and $f$ and $i$ are flavour indices\footnote{Note that in addition to the terms in \eq{Yukawa}, one could have a term $\lambda _i^S{{\bar D}_L}{d_i} S^\dag$, which, however, without loss of generality, can be removed by rotations of $d_i,\,D_R$ and an appropriate redefinition of the couplings. }. Here we choose to work in the down basis where $Y^d$ is diagonal and the CKM matrix originates from the up-sector. In addition, there is only one possible coupling of the charged scalar to leptons allowed by our charge assignment
$
- {\cal L}_Y^{L\phi^+} = \xi \bar L_2^c \cdot L_1\phi^+ +{\rm h.c.}\,
$ were
$\cdot$ stands for a contraction in $SU(2)_L$ space via the anti-symmetry tensor. 


The vacuum expectation value $\langle S\rangle={v_{S}}/{\sqrt{2}}$ generates masses for the $Z^\prime$ boson ($m_{Z^\prime}=g^\prime v_S$ ) the VLQs ($m_Q=v_S/\sqrt{2}\eta_Q$, $m_D=v_S/\sqrt{2}\eta_D$) as well as the mass of the charged and neutral singlet ($m_{\phi}={\mathcal O}(v_S)$, $m_S={\mathcal O}(v_S)$). After EW symmetry breaking, the quark doublet is decomposed into its components and CKM rotations generate ${V_{fj}}\kappa _j^\phi {\mkern 1mu} {{\bar u}_{Lf}}{\mkern 1mu} {\phi ^ + }{Q_R}$ couplings which will later be relevant for $D^0-\bar D^0$ mixing.

\section{Explaining the Anomalies}
\label{observables}

\begin{figure*}[t]
	\includegraphics[width=0.9\linewidth]{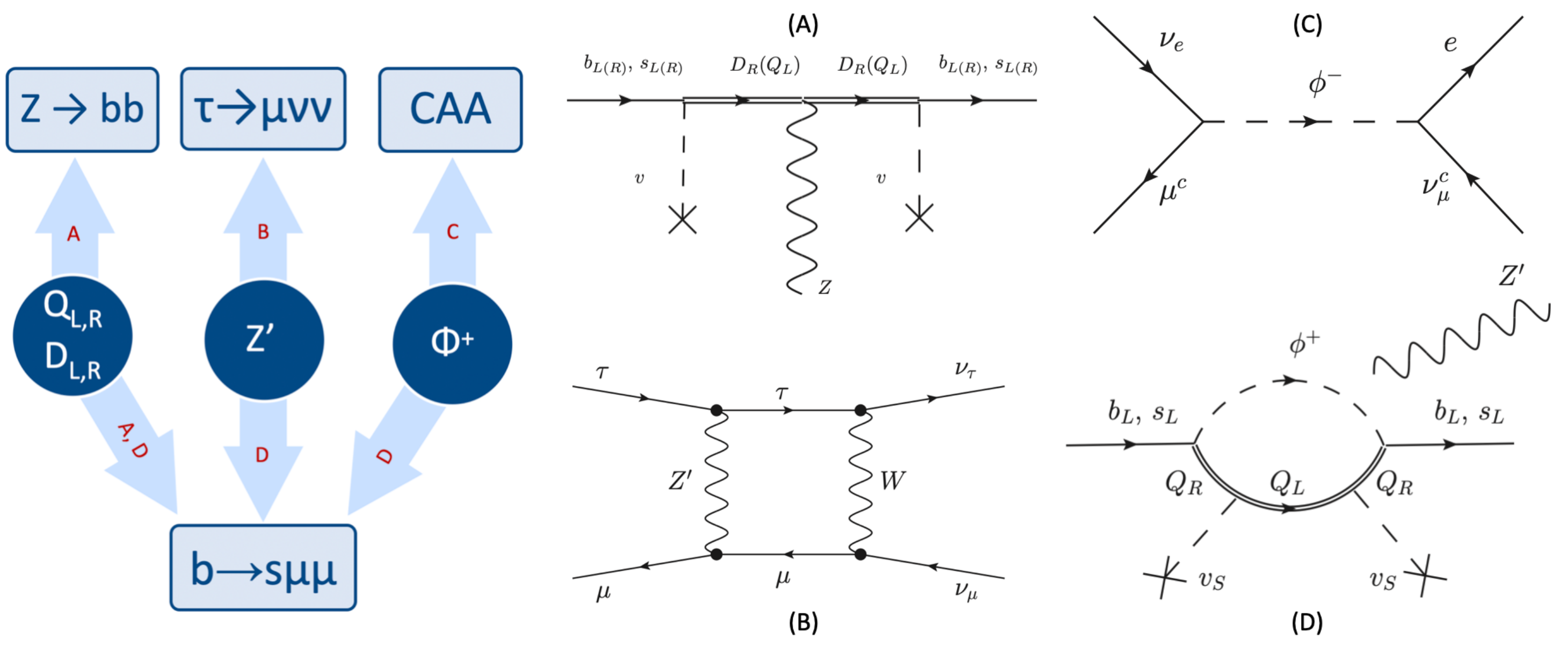}
	\caption{Diagramatic representation of how the Feynman diagrams (A)-(D) within our model contribute to $Z\to \bar{b}b(\bar{s}s)$, muon decay, $\tau\to\mu\nu\nu$ and $b\to s\ell\ell$ and explain the associated anomalies.\label{AnomaliesPlot}}
\end{figure*}

Let us now discuss how our model can explain the anomalies and which observables are relevant in constraining it, starting with $Z b\bar b$.

\begin{boldmath}
	\subsection{$Z b\bar b$}
\end{boldmath}

The mixing of the heavy quarks with the SM ones leads to desired modifications of the $Zbb$ couplings via diagram (A) shown in Fig.~\ref{AnomaliesPlot}. Using the publicly available HEPfit code~\cite{deBlas:2019okz}, we updated the fit of Refs.~\cite{deBlas:2016ojx,deBlas:2016nqo}, finding
\begin{align}
\begin{split}
\left| {\lambda _b^Q} \right|^2 \!\!\!=\! (1.12\pm 0.46)\dfrac{{M_Q^{}}}{{ \rm TeV}},\,
\left| {\lambda _b^D} \right|^2 \!\!\!=\! (0.18\pm 0.09)\dfrac{{M_Q^{}}}{{ \rm TeV}}.
\end{split}
\end{align}
Note that this combination of couplings is not significantly constrained from other observables so that we can fully account for the anomaly.

\begin{boldmath}
	\subsection{Cabibbo Angle Anomaly}
\end{boldmath}

The Cabibbo Angle Anomaly originates from a (apparent) deficit in $1^{\rm st}$ row CKM unitarity. Equivalently, it manifests itself in a disagreement between the determinations of $V_{us}$ from kaon and tau decays vs $V_{us}$ from super-allowed beta decays (assuming CKM unitarity). Following Ref.~\cite{Crivellin:2020ebi} we have
\begin{align}
V_{us}^\beta =0.2281(7)\,,\;\;V_{us}^\beta|_{\text{NNC}} =0.2280(14)\,,
\label{Vusbeta}
\end{align}
where the latter value contains the  ``new nuclear corrections'' (NNCs) proposed in Refs.~\cite{Seng:2018qru,Gorchtein:2018fxl}. Since at the moment the issue of the NNCs is not settled, we will perform our fit for both determinations. The value of $V_{us}^\beta$ has to be compared to $V_{us}$ from kaon~\cite{Aoki:2019cca} and tau decays~\cite{Amhis:2019ckw} 
\begin{align}
\begin{split}
V_{us}^{K_{\mu 3}}&=0.22345(67)\,,\;\;\; V_{us}^{K_{e 3}}=0.22320(61)\,,\\
V_{us}^{K_{\mu 2}}&=0.22534(42)\,,\;\;\;V_{us}^\tau = 0.2195(19)\,,
\end{split}
\label{VusKl3}
\end{align}
which are significantly lower.


The Feynman diagram (C) in Fig.~\ref{AnomaliesPlot} generates a necessarily constructive effect w.r.t the SM in $\mu\to e\nu\nu$ and modifies the determination of the Fermi constant ($G_F$) from muon decays.
 While the $V_{us}$ determination from kaon and tau decays is mostly independent of the Fermi constant, $V_{us}$ from super-allowed beta decays even has a sensitivity enhanced by $V_{ud}^2/V_{us}^2$~\cite{Crivellin:2020lzu}. This means that the ``real'' Lagrangian value of $V_{us}$ of the unitary CKM matrix in terms of the one measured from beta decays is given by
\begin{equation}
V_{us}^{} = V_{us}^\beta \left( {1 - \left| {\frac{{V_{ud}^2}}{{V_{us}^2}}} \right|\left| \frac{{\xi ^2}}{{g_2^2}} \right|\frac{{m_W ^2}}{{m_\phi^2}}} \right)\,.
\end{equation}

 As a modification of $G_F$ also affects the EW sector we included the determinations of $V_{us}$ in \eq{Vusbeta} and \eq{VusKl3} into HEPfit and performed a global fit finding
\begin{align}
\begin{split}
V_{us}^\beta: & \left| \xi  \right|^2 = \left( {0.043\pm 0.010} \right)\frac{{m_\phi ^{2}}}{{\rm TeV^2}}\,,
\\
V_{us}^\beta|_{\text{NNC}}:& \left| \xi  \right|^2 = \left( {0.021 \pm 0.013} \right)\frac{{m_\phi ^{2}}}{{\rm TeV^2}}\,.
\end{split}
\end{align} 
Note that this range for $\xi$ brings $V_{us}$ from beta decays into agreement with $V_{us}$ from $K\to\mu\nu/\pi\to\mu\nu$\footnote{There is a small correlations between the NP effect in $G_F$ and $Z\to b\bar b$ such that for the central value of $\xi/m_\phi$ only a smaller NP effect in left-handed $Zbb$ coupling is needed.}. Therefore, also with respect to the CAA we improve by $\approx 2\,\sigma$ w.r.t the SM, depending on the value of $V_{us}^\beta$ considered.

\begin{boldmath}
\subsection{$\tau\to\mu\nu\nu$}
\end{boldmath}

Let us now study the effect of the $Z^\prime$ in $\tau\to\mu\nu\nu$ which is modified by diagram (B) in Fig.~\cite{Altmannshofer:2014cfa}, resulting in~\cite{AnomaliesPlot}
\begin{equation}
\frac{\operatorname{BR}\left(\tau \rightarrow \mu \nu \bar{\nu}\right)}{\operatorname{BR}\left(\tau \rightarrow e \nu \bar{\nu}\right)_{\operatorname{SM}}}=\frac{\operatorname{BR}\left(\tau \rightarrow \mu \nu \bar{\nu}\right)}{\operatorname{BR}\left(\mu \rightarrow e \nu \bar{\nu}\right)_{\operatorname{SM}}} \simeq 1+\Delta\,,
\label{taumununu}
\end{equation}
with
\begin{align}
\Delta=\frac{3\left(g^{\prime}\right)^{2}}{4 \pi^{2}} \frac{\log \left(m_{W}^{2} / m_{Z^{\prime}}^{2}\right)}{1-m_{Z^{\prime}}^{2} / m_{W}^{2}}=(4.7\pm 2.3)\times 10^{-3}\,.
\end{align}
The experimental value is obtained by averaging the measurements of both ratios, including the correlation of 0.48~\cite{Amhis:2019ckw}. In particular, for $m_{Z^\prime}=1\,$TeV we find
\begin{equation}
(g^\prime)^2=(1.9\pm 0.9)\,\frac{m_{Z^\prime}}{{\rm TeV}}\,,
\end{equation}
neglecting logarithmic effects. Notice that the $Z^\prime$ only affects the numerator of these ratios while $\mu\to e\nu\nu$ is affected by tree-level $\phi^+$ exchange as discussed above. However, the latter effect is stringently bounded by $V_{us}$ and the EW fit such that its impact on \eq{taumununu} is negligible.

The $Z^\prime$ also contributes to neutrino trident production (NTP)~\cite{Altmannshofer:2014pba,Geiregat:1990gz,Mishra:1991bv,Adams:1998yf} and gives rise to the loop-corrections of $Z$ couplings to charged leptons and neutrinos~\cite{Altmannshofer:2014cfa,Haisch:2011up}. However, neither these effect nor NTP are in conflict with the preferred region from $\tau\to\mu\nu\nu/\tau\to e\nu\nu$ and $\tau\to\mu\nu\nu/\mu\to e\nu\nu$. Therefore, we can obtain the best fit point for ${\rm Br}[\tau\to\mu\nu\nu]/{\rm Br}[\tau\to e\nu\nu]$ and ${\rm Br}[\tau\to\mu\nu\nu]/{\rm Br}[\mu\to e\nu\nu]$ and thus improve on the SM by more than $2\,\sigma$.
\bigskip

\begin{boldmath}
\subsection{$b\to s\ell^+\ell^-$} 
\end{boldmath}

Here we want to explain $b\to s\ell^+\ell^-$ data with a combination of a LFU effect from modified $Zsb$ coupling (diagram (D)) and a LFU violating one originating from the $Z^\prime$ (diagram (A)). The former one is generated at tree-level via the mixing of the 
vector-like quark with SM quarks and it is given by
\begin{align}
\dfrac{{\cal C}_{9^{(\prime)}}^{\rm U}}{1-4s_w^2} =-{\cal C}_{10^{(\prime)}}^{\rm U}=  \frac{{2\pi^2}}{{e^2M_Q^2}}\frac{\lambda _s^{Q(D*)}\lambda_b^{Q*(D)}}{{\sqrt 2 {G_F}{V_{tb}}V_{ts}^*}}\,,
\end{align}
in the conventions of Ref.~\cite{Alguero:2018nvb}. For the latter one, the $Z^\prime sb$ coupling is generated at the loop level through diagrams like the one shown in Fig.~\ref{AnomaliesPlot}. We parameterize the effective $Z^\prime$ coupling to down-quarks generically as
\begin{align}
\begin{aligned}
{\cal L} &= {\bar d_f}{\gamma ^\mu }\left( {\Delta _{fi}^{\prime dL}{P_L} + \Delta _{fi}^{ \prime dR}{P_R}} \right)Z_\mu ^{ \prime}{d_i}\,,
\end{aligned}
\end{align}
and obtain
\begin{equation}
\Delta_{fi}^{\prime dL} = g^\prime\frac{{\kappa _f^\phi \kappa _i^{\phi *}}}{{16{\pi ^2}}}\frac{{ - x + x\log (x) + 1}}{{{{(x - 1)}^2}}}\,,\label{Zpdd}
\end{equation}
with $x = {{m_\phi ^2}}/{{M_Q^2}}$. 
This results in the purely LFU violating effects
\begin{align}
\begin{aligned}
{{\cal C}_{9\mu}^{\rm V}}=-\frac{16\pi^2}{e^2}\frac{\Delta^{\prime dL}_{23}(\Delta^{\prime\ell R}_{22}+\Delta^{\prime\ell L}_{22})}{4\sqrt{2}G_F M_{Z^\prime}^2 V_{tb}V_{ts}^*}\,.
\end{aligned}
\end{align}


Performing a global fit within this scenario\footnote{Note that in the simpler case with only the VLQ $Q$ (but not $D$) we would obtain to a good approximation scenario 11 of Ref.~\cite{Alguero:2019ptt} with a pull of $6.3\,\sigma$.} we find a pull of $6.1\,\sigma$ w.r.t the SM and a p-value of 47.5\%. The best fit points and $1\,\sigma$ CL intervals for the Wilson coefficients are 
\begin{align}
\begin{split}
{\cal C}^{\rm V}_{9\mu}&=-1.06 \pm 0.16 \,,\\
{\cal C}^{\rm U}_{10^{\prime}} &= -0.24 \pm 0.17\,, \; {\cal C}^{\rm U}_{10}= 0.18\pm 0.19\,.
\end{split}
\end{align}

However, $b\to s\ell^+\ell^-$ cannot be explained without affecting $\Delta F=2$ processes. While tree-level $Z$ and $Z^{\prime}$ effects turn out to be negligible (due to the tiny $sb$ couplings) $\phi^+$ box contributions generate
\begin{align}
{{\cal C}_1^{\phi^+} = \dfrac{(\kappa_s^\phi\kappa_b^{\phi*})^2}{128\pi^2}\frac{m_\phi^4-2m_\phi^2m_Q^2\log{\frac{m_\phi^2}{m_Q^2}}-m_Q^4}{(m_\phi^2-m_Q^2)^3}}\,,
	\label{DeltaFHMatching}
\end{align}
following the conventions in Ref.~\cite{Ciuchini:1998ix}. Including the 2-loop RGE of Ref.~\cite{Ciuchini:1997bw,Buras:2000if} and the bag factor of Ref.~\cite{Aoki:2019cca} we find at the $B_s$ meson scale
\begin{align}
\frac{{\Delta {m_{{B_s}}}}}{{\Delta m_{{B_s}}^{\rm SM}}}= 1 + 1.1\, {{{\cal C}_1^{\phi^+}}}\!\times \! {10^{10}}{\rm GeV}{^2}=1.11\pm0.09\,,
\end{align}
for real NP contributions according to the global fit of Ref.~\cite{Bona:2007vi}.
Similarly, we get a bound from CP violation in the $D^0-\bar D^0$ system~\cite{Bevan:2014tha,Bazavov:2017lyh} 
\begin{equation}
2.3| {\rm Im}[{\cal C}_1^{D^0-\bar D^0}]|\!\times \!10^{12}{\rm GeV}^2<0.033\,,
\end{equation}
with ${\cal C}_1^{D^0-\bar D^0}$ obtained from \eq{DeltaFHMatching} by replacing $\kappa_s^\phi\kappa_b^{\phi*}$ with $V_{us}\kappa_s^\phi (V_{cs}^*\kappa_s^{\phi*}+V_{cb}^*\kappa_b^{\phi*})$. For the experimental limit we assumed that the SM contribution to CP violation in $D^0-\bar D^0$ mixing is negligible.

\begin{figure}[t!]
	\includegraphics[width=0.95\linewidth]{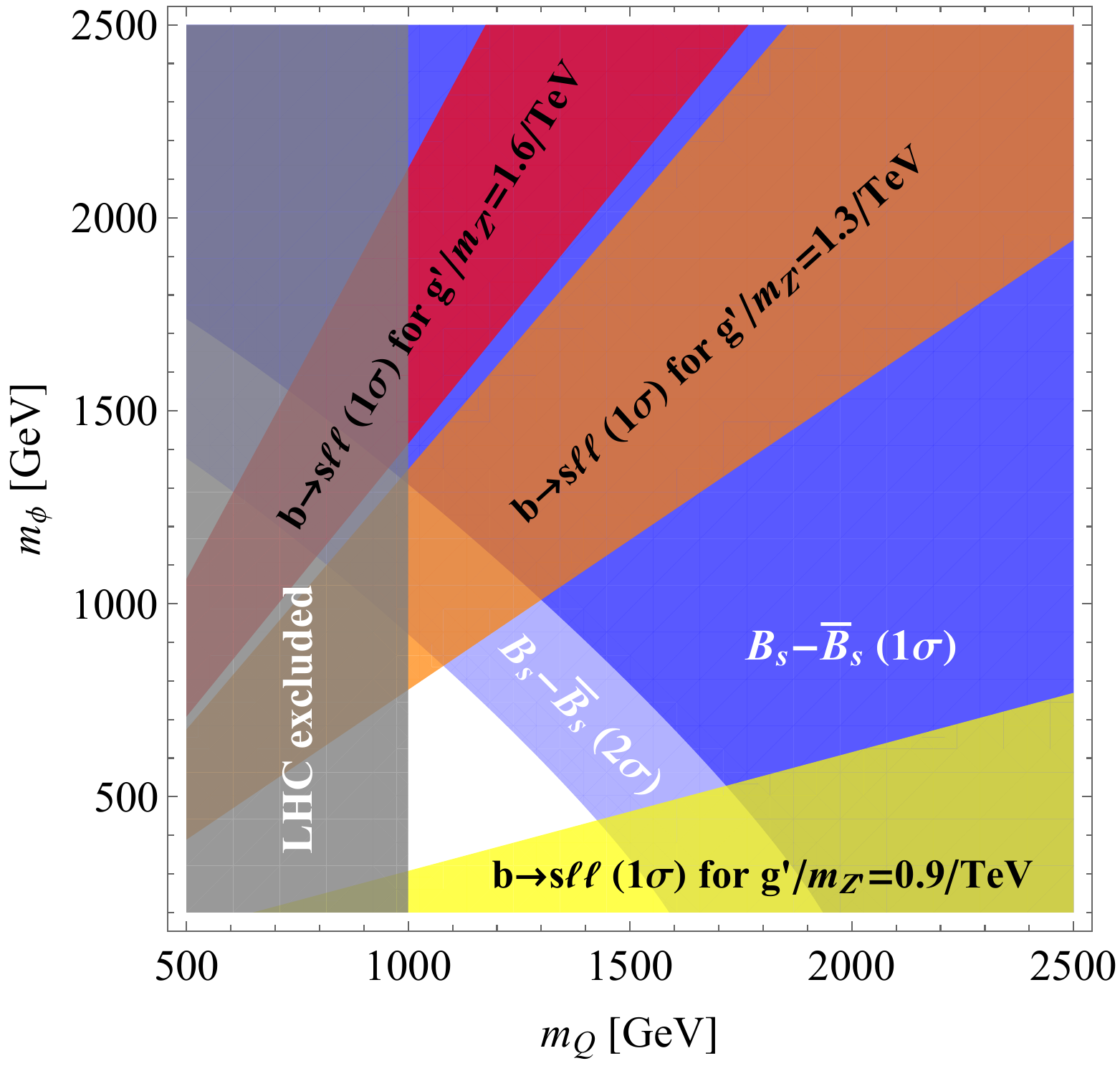}
	\caption{Preferred and excluded regions in the $m_Q$-$m_\phi$ plane for $\kappa_s\kappa_s^*=-0.3$ and $m_Q=m_D$. Note that for $m_Qm_\phi>1.5{\rm TeV}^2$ one can account for $b\to s\ell^+\ell^-$ data while being in agreement with $B_s-\bar B_s$ mixing at the $1\,\sigma$ level.\label{bsllBsmixing}}
\end{figure}

Turning to the phenomenological analysis, notice that we can generate ${\cal C}_{10^{(\prime)}}^{\rm U}$ from the modified $Zbs$ couplings without generating a relevant effect in $B_s-\bar B_s$ mixing. Therefore, we can account for the full range of values for ${\cal C}_{10^{(\prime)}}^{\rm U}$ preferred by the fit to $b\to s\ell^+\ell^-$ data. Similarly, note that generating ${\cal C}_{9\mu}^{\rm V}$ from the $Z^\prime$ penguins also gives rise to an effect in $D^0-\bar D^0$ mixing due to CKM rotations. However, the resulting constraint is sub-leading for $\kappa_b>\kappa_s$. Therefore, we find the results shown in Fig.~\ref{bsllBsmixing} from which it is clear that we can reach the best fit point for $b\to s\ell^+\ell^-$ without being in conflict with $B_s-\bar B_s$ mixing while choosing $g^\prime/m_{Z^\prime}$ as preferred by ${\rm Br}[\tau\to\mu\nu\nu]/{\rm Br}[\tau\to e\nu\nu]$ and ${\rm Br}[\tau\to\mu\nu\nu]/{\rm Br}[\mu\to e\nu\nu]$. Note that LHC bounds are not important here due to the small couplings to quarks. Therefore, we can improve the fit compared to the SM by $6.1\,\sigma$ in the $b\to s\ell^+\ell^-$ alone\footnote{Note that recasting the ATLAS analysis~\cite{Aad:2019fac} for our $Z^\prime$ we find that bounds are not constraining since the couplings to quarks are not only loop-induced and therefore small but also the production cross section is reduced by the small bottom PDF.}. 
\bigskip

\section{Conclusions and Outlook}\label{conclusions}

In this article we proposed a simple model obtained from the SM by adding: 
	\vspace{-2mm}
\begin{itemize}
	\item Two heavy quarks which are vector-like ($Q$ and $D$) under the SM gauge group.
	\vspace{-2mm}
	\item A gauged $L_\mu-L_\tau$ symmetry resulting in a $Z^\prime$ boson.
	\vspace{-5mm}
	\item A neutral and a singly charged scalar, singlet under color and weak isospin, ($S$ and $\phi^+$) with $L_\mu-L_\tau$ charge $-1$.
\end{itemize}
	\vspace{-1mm}
This model can explain:
	\vspace{-1mm}
\begin{itemize}
	\item The $Z\to b\bar b$ forward-backward asymmetry via the mixing of the 
	vector-like quarks with the SM bottom quark.
		\vspace{-2mm}
	\item The 
	Cabibbo Angle Anomaly via a positive definite shift in $G_F$ induced by the singly charged scalar.
			\vspace{-2mm}
	\item $\tau\to\mu\nu\nu/\tau\to e\nu\nu$ and $\tau\to\mu\nu\nu/\mu\to e\nu\nu$ via the box contributions involving the $Z^\prime$.
		\vspace{-2mm}
	\item Accounts for $b\to s\ell^+\ell^-$ data through a modified $Z$ coupling and a loop induced $Z^\prime$ effect without being in conflict with $B_s-\bar B_s$ mixing.
\end{itemize}
This is illustrated in Fig.~\ref{AnomaliesPlot}.

Therefore, our model describes data significantly better than the SM and constitutes the first unified explanation of all four anomalies. With new particles at the TeV scale, it provides interesting discovery potential for the (HE-) LHC~\cite{Abada:2019ono} and the FCC-hh~\cite{Benedikt:2018csr} but could also be indirectly verified through $Z$ pole observables by FCC-ee~\cite{Abada:2019zxq}, ILC~\cite{Baer:2013cma}, CEPC~\cite{An:2018dwb} or CLIC~\cite{Aicheler:2012bya}. Also BELLE II is sensitive to the $Z\to b\bar b$ asymmetry via $e^+e^-\to b\bar b$ measurements with polarized electron beams~\cite{Roney:2019til}. Furthermore, precision measurements of $\tau$ decays at BELLE~II~\cite{Kou:2018nap} and of course the pattern predicted in $b\to s\ell^+\ell^-$ at the HL-LHC~\cite{Cerri:2018ypt} and BELLE~II~\cite{Kou:2018nap} could test our model. 

In this work we presented the minimal model capable of providing an explanation of the hints for NP. As it possesses gauge anomalies, it is interesting to look for extensions which are free of this obstacle (see e.g. Refs.~\cite{Ellis:2017nrp,Aebischer:2019blw} for accounts in the context of $b\to s\ell\ell$). For example, by adding two more heavy quarks $Q^\prime_{L,R}$ and $D^\prime_{L,R}$ with the same representation under $SU(3)\times SU(2)_L$ as $Q_{L,R}$ and $D_{L,R}$ but with opposite $U(1)_Y$ and $U(1)^\prime$ charge this can be easily achieved without any significant effect on the phenomenology. In general, embedding our model into a more unified framework would be very interesting and opens up novel avenues for model building. 
\medskip

\begin{acknowledgements}
	The work of A.C. is supported by a Professorship Grant (PP00P2\_176884) of the Swiss National Science Foundation.
	JM  gratefully acknowledges the financial support by ICREA under the ICREA Academia programme. JM and MA received financial support from 	Spanish Ministry of Science, Innovation and Universities (project FPA2017-86989- P) and from the Research Grant Agency of the Government of Catalonia (project SGR 1069).	 
\end{acknowledgements}

\newpage

\bibliography{bibliography}

\end{document}